\documentclass{aa}
\usepackage[comma,authoryear]{natbib}
\usepackage{graphics}
\usepackage[latin1]{inputenc} 

\begin{document}

\title{Comparison of different exoplanet mass detection limit methods using a sample of main-sequence intermediate-type stars}

   \titlerunning{Estimation of exoplanets detection limits}

   \author{N. Meunier \inst{1}, A.-M. Lagrange \inst{1}, K. De Bondt\inst{1} 
	  }
   \authorrunning{Meunier et al.}

   \institute{
UJF-Grenoble 1 / CNRS-INSU, Institut de Plan\'etologie et d'Astrophysique de Grenoble (IPAG) UMR 5274, Grenoble, F-38041, France\\
  \email{nadege.meunier@obs.ujf-grenoble.fr}
             }

\offprints{N. Meunier}

   \date{5 March 2012 ; 13 June 2012}

\abstract{The radial velocity (RV) technique is a powerful tool for detecting extrasolar planets and deriving mass detection limits that are useful for constraining 
planet pulsations and formation models.}
{Detection limit methods must take into account the temporal distribution of power of various origins in the stellar signal. These methods 
must also be able to be applied to large samples of stellar RV time series}
{We describe new methods for providing detection limits. We compute the detection limits for a sample of ten main sequence stars, which are of G-F-A type, in general active, and/or 
with detected planets, and various properties. We 
use them to compare the performances of these methods with those of two other methods used in the litterature. }
{We obtained detection limits in the 2-1000 day period range for ten stars. Two of the proposed methods, based on the correlation between 
periodograms and the power in the periodogram of the RV time series in specific period ranges, are robust and represent a significant improvement 
compared to a method based on the root mean square of the RV signal. }
{We conclude that two of the new methods (correlation-based method and local power analysis, i.e. LPA, method) provide robust detection limits, which are better 
than those provided by methods that do not take into account the temporal sampling.}

\keywords{Techniques: radial velocities -- Stars: planetary systems -- Stars: early-type -- Stars: individual (HD 10180, HD 60532, HD 105690, HD 115892, HD 124850, HD 172555, HD 199260, HD 210302, HD 219482, $\beta$ Pic)}

\maketitle

\section{Introduction}

The radial velocity (hereafter RV) technique is a powerful tool for detecting planets, but also for deriving detection limits 
(i.e. the upper limit to possible planet masses for different periods). Detection limits indeed represent invaluable information for either  
the study of specific objects, or and above all the derivation of quantitative constraints on the formation processes of planets. This is true for 
detection limits obtained with RV as well as direct imaging. Different criteria have been used to compute these detection limits. One is based 
on the root mean square (rms) of the RV data compared to the rms of the planetary RV \cite[][for the principle]{galland05} and has been used by \cite{lagrange09}.  

This rms-based method is very fast, but can significantly  overestimate the detection limit in some cases. The signal of the
planet (which has a certain period) is compared to the rms of the whole signal, which may contain strong power at periods very different
from the planet period. This was the case for \object{$\beta$ Pic} \cite[][hereafter paper I]{lagrange12}: \object{$\beta$ Pic} is a pulsating star, with a
strong power in the domain 20-30
minutes, which dominates the rms computed over the RV signal, but this power is much weaker in the frequency
domain in which we search for planets. Detection limits based on this rms are then overestimated. We therefore derived other methods that allow us to take into account the temporal behavior of the stellar noise. 
In paper I, we presented the first results for \object{$\beta$ Pic} and showed that we could improve the detection limits significantly for periods in the range from a few days to 
a few hundreds days. Here, we present these methods in detail and use them on a sample of ten stars (including
\object{$\beta$ Pic} for comparison purposes) with various characteristics. We test their robustness, as the detection limit depends on the available data 
(temporal sampling) and the temporal structure of the stellar noise. 
These tests are made on a limited number of stars, that are representative of our large (250 stars) sample covering either early-type (A, F) stars or young solar-type stars. We also added for comparison purposes a slowly rotating solar-type main-sequence (MS) star. Their v.$\sin$i ranges from a few km/s up to 17 km/s. 
Our long-term goal is to use these methods 
to obtain detection limits for the $\sim$250 stars of our complete set of MS A-F stars, some of which are young stars sample 
that have been surveyed in the northern and southern hemispheres during searches for planets. 
Given the characteristics of our stellar sample, we focus mainly on Jupiter-mass planets.

The star sample is described in Sect.~2. 
We compute the detection limits using four different methods: rms, correlation, peak, and {\it local power amplitude} LPA, respectively 
described in Sect.~3 to 6.  In Sect.~3 to 6, we also present the results and test the robustness
of each of these methods for each of the relevant parameters. In Sect.~7, the detection limits obtained 
with the different methods are discussed and compared to each other, and we discuss specific stellar cases. We present our conclusions in Sect.~8.

\section{Star sample}

\begin{figure*}
\includegraphics{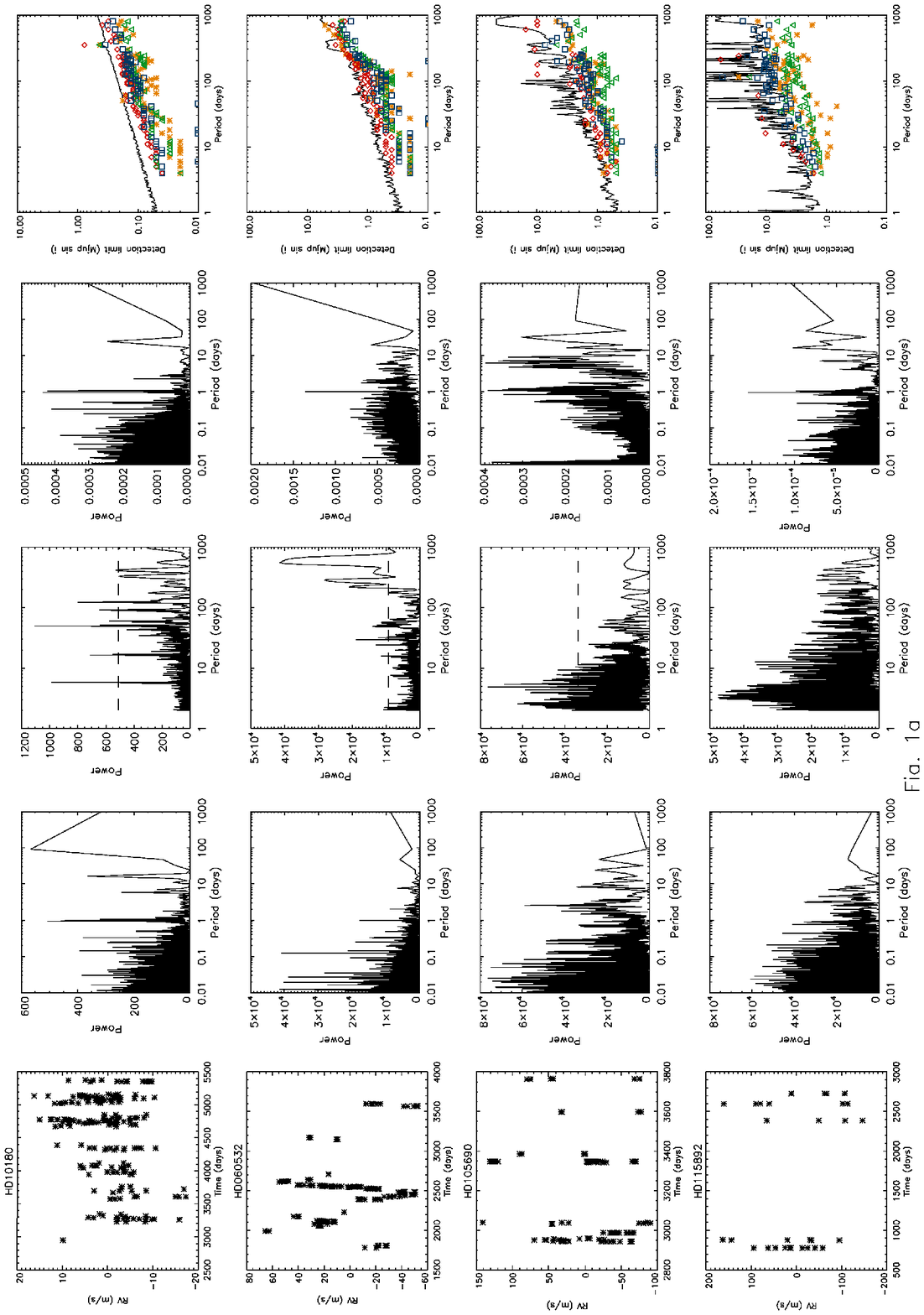}
\caption{{\it First column}: RV versus time (JD--2452000 days) for each star of the sample. 
{\it Second column}: corresponding periodogram over the range 0.01--1000 days. 
{\it Third column}: Same for the 2--1000 day range. 
{\it Fourth column}: periodogram of the temporal sampling. 
{\it Fifth column}: corresponding detection limit versus the period, for the correlation-based method (stars, orange), for the peak method (diamonds, red), for the rms method (solid line, black), for the LPA method (triangle, green), and for the bootstrap method (squares, blue).}
\label{res_a}
\end{figure*}
\begin{figure*}
\includegraphics{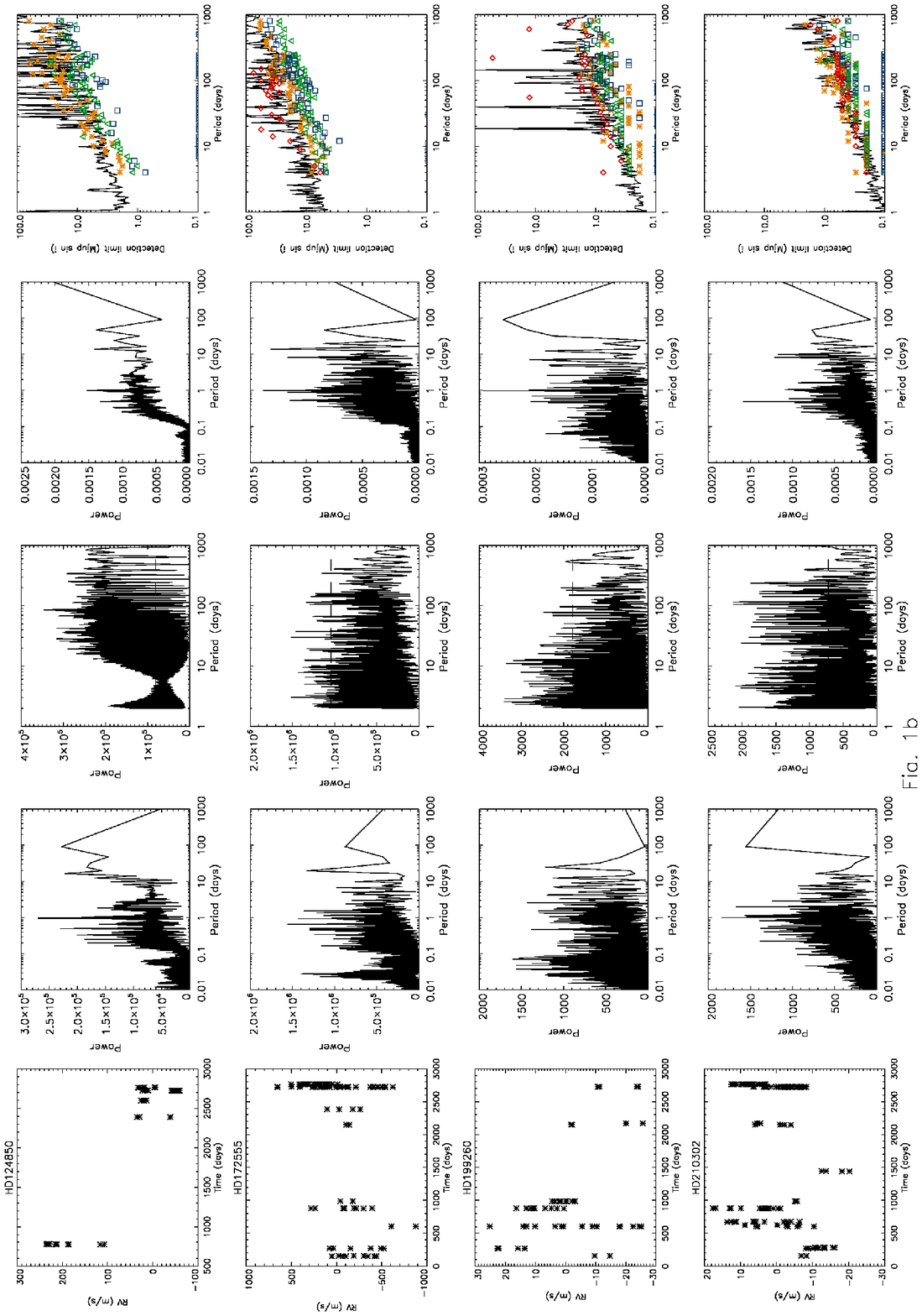}
\end{figure*}
\begin{figure*}
\includegraphics{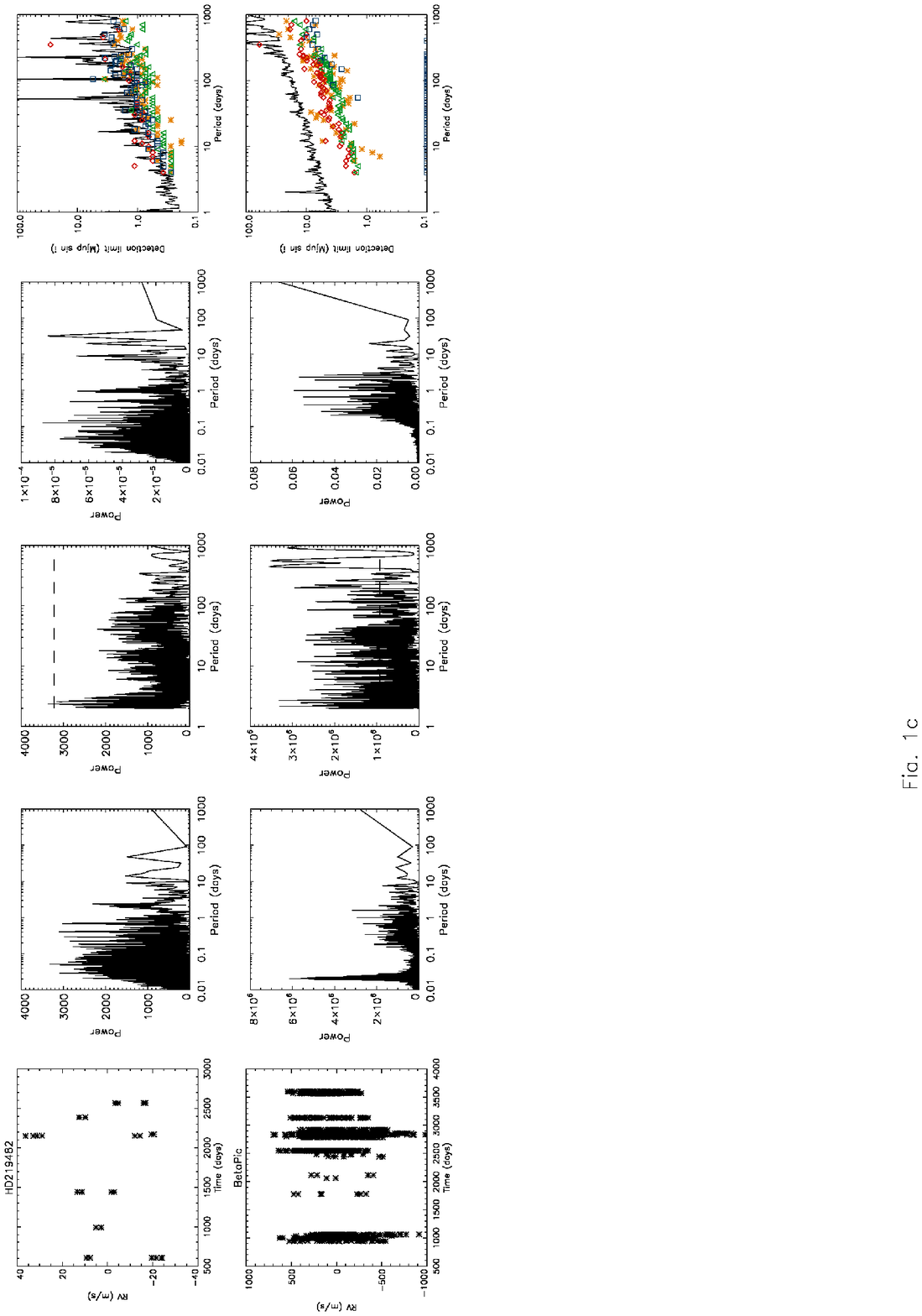}
\end{figure*}

Our sample is made of 10 stars, most of them observed in the framework of our survey described above. Their 
properties are listed in Tab.~\ref{tabstar}. They were chosen to span various conditions in term of observation sampling
and stellar properties (such as pulsations, activity or planet presence, and v.$\sin$i). For example, \object{HD60532} 
is a star with two planets, with 201 day and 604 day periods \cite[]{desort08}.
\object{HD10180} has the smallest rms RV of our sample, and the RV curve could be fitted by up to seven planets by \cite{lovis11a}. In the following, we
mostly consider the planets with 5.8 day and 49.7 day periods (with masses of, respectively, 13.2 and 25.4 M$_{\rm Earth}$), as their signals appear to 
account for  the largest part of the rms RV. 
With more than 1000 data points, \object{$\beta$ Pic} has by far been the most well-observed, while on the other hand \object{HD219482} has been chosen 
because it has been poorly observed (only 24 data points).
\object{HD105690} has a good temporal sampling, although over a short duration compared to the other stars (about three years). Two stars display pulsations
(\object{HD172555} and \object{$\beta$ Pic}). The rms RV is the largest for these two pulsating stars. 
The RV time series of these stars are shown in Fig.~\ref{res_a} (first column), as well as the corresponding periodogram over both a wide range of periods (0.01--1000 days, second column) and a smaller range (2-1000, third column), which is  used in the next few sections. The fourth column provides the periodogram of the temporal sampling for comparison.  
In the following, all detection limits are computed for planets with no eccentricity. All masses correspond to the planet mass times the v.$\sin$i.

\begin{table*}
\begin{center}
\caption{Star sample and characteristics }
\label{tabstar}
\begin{tabular}{ccccccccccc}
\hline \hline
\# & Star        & Type & v.$\sin$i & $B-V$ & Mass          & Nb of obs. & Obs. length & RV rms & Activity & Sampling \\ 
   &             &      & (km/s)    &       & (M$_{\odot}$)    &            & (days)      & (m/s)  &  &  \\ \hline
1 & \object{HD10180}   & G1V    & $<$3       & 0.63 &   1.06        &   190      &   2428       &   6.3   & planets   &  large data set \\
2 & \object{HD60532}   & F6IV-V & 10 & 0.48  & 1.38          & 164        & 1822       &27.6   & 2 planets     & large data set \\
3 & \object{HD105690}   & G5V   &  9.6         & 0.71  & 0.88          & 104        & 824         &57.1  & low      & average data set \\
4 & \object{HD115892}   & A2V   &   90        & 0.07  & 2.36          & 30     & 1953        &84.5  & low      & small data set \\
5 & \object{HD124850}   & F5V   &   15        & 0.51  & 1.36          & 95         & 1992        &88.4  &  RV trend     & av. data set, 2 packs \\
6 & \object{HD172555}   & A7V   &   175        & 0.20  & 1.83          & 99         & 2619        &306.6 & pulsations       & average data set \\
7 & \object{HD199260}   & F7V   &    13       & 0.51  & 1.36          & 51         & 2571        &13.9  & low      & average data set \\
8 & \object{HD210302}   & F6V   &     12      & 0.49  & 1.37          & 128        & 2610        &7.9    & low    & large data set \\
9 & \object{HD219482}   & F7V   &     7      & 0.52  & 1.36          & 26         & 1966        &18.9   & low     & small data set \\
10 & \object{$\beta$ Pic} & A6V   &    115       & 0.17 & 1.75          & 1049       & 2658        &275.4   & pulsations     & very large data set \\ \hline
\end{tabular}
\end{center}
\end{table*}


\section{Root-mean-square-based method}

\subsection{Method and results}

The principles  behind the rms-based method are described in \cite{galland05}. It is used as a reference
in this paper, hence we describe it briefly and test it in a similar way. This method is based on the comparison of the measured RV jitter (rms RV of the data) with the rms of the RV that would be expected for a planet observed at the same dates. For a given planet mass and period, several realizations of the planet signal (over the observed sampling) are computed, each realization 
corresponding to a different phase of the planet. The rms RV for each realization is computed and compared to the observed RV rms. If all values are above the observed rms RV, then the mass
is above the detection limit. We typically use 1000 realizations: the derived detection limit means that the presence of a planet with a higher mass is possible with a  probability smaller than 1/1000.

The results are shown in Fig.~\ref{res_a} (last column, solid line). 
Since for all planet periods the signal is compared to a single rms RV, the detection 
limit would follow a straight line in a log-log plot if the data were perfectly sampled in time. 
The few peaks with larger detection limits are produced by the temporal window of the observations. 

\subsection{Impact of parameters}

This method has the advantage of using very few parameters. The only parameter is indeed the number of realizations over which the computation is made,
i.e. the probability of having a planet with a mass higher than the detection limit. We therefore compared the detection limits computed for probabilities of 1/100 and 1/1000, for the 60 planet periods computed for the other methods. As shown in Table.~\ref{tabres}, the results is the same for all stars and periods.  

\subsection{Conclusion on the rms-based method}

The rms-based method is both  fast and robust. However, we know that it may significantly overestimate the detection limit for some periods as the 
temporal structure of the observed signal is not taken into account. This justifies out trial of  the methods studied in the next few sections.

\section{Correlation-based method}

\subsection{Method and results}

\begin{figure}
\includegraphics{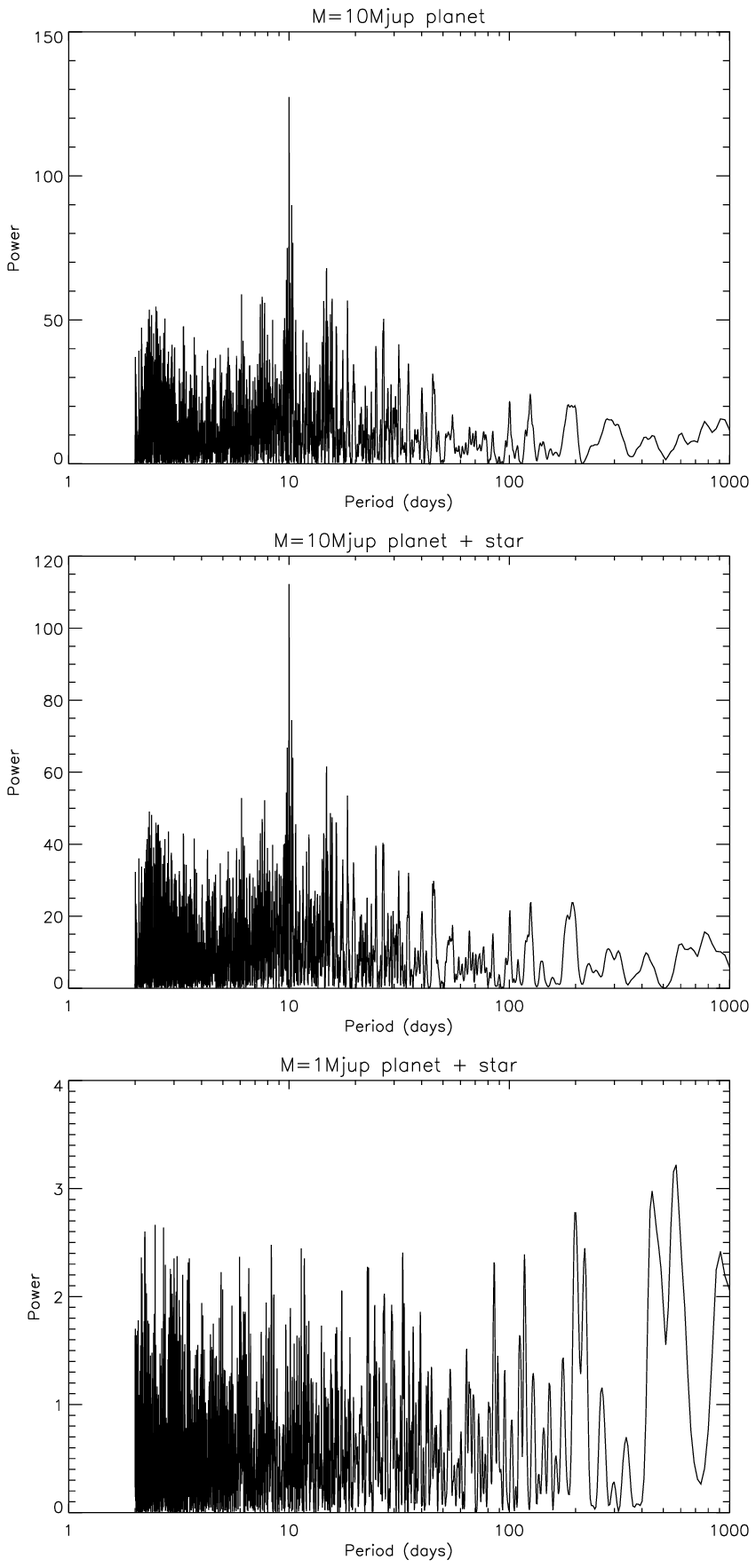}
\caption{{\it Upper panel}: periodogram of a 10 Mjup planet RV signal at $P$=10 days using the \object{$\beta$ Pic} temporal sampling. {\it Middle panel}: same for a 10 Mjup planet added to the observed star RV. {\it Lower panel}: same for a 1 Mjup planet added to the star RV.}
\label{period}
\end{figure}

\begin{figure}
\includegraphics{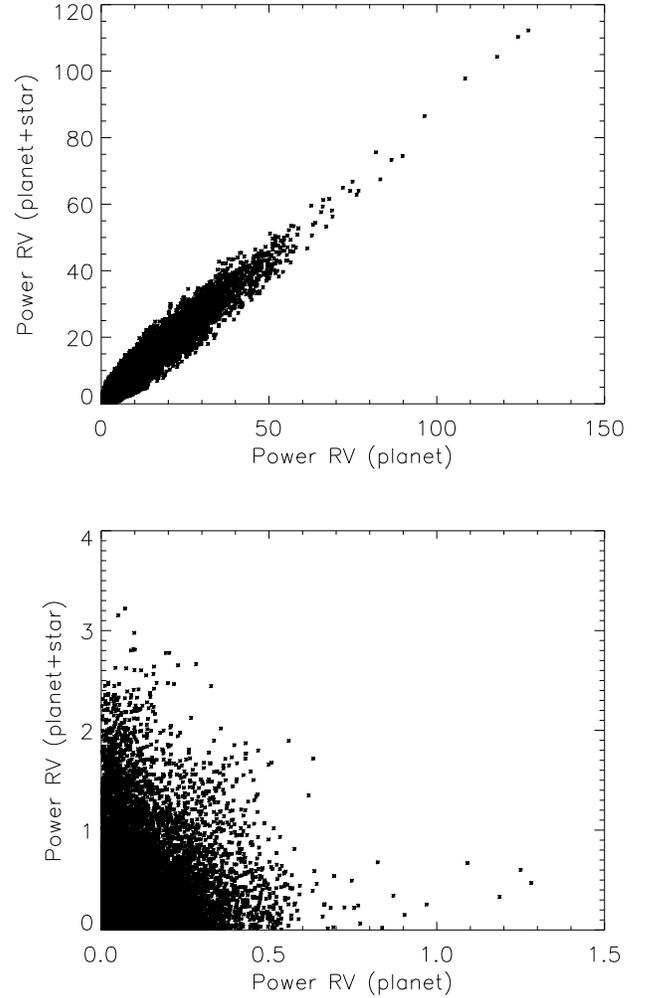}
\caption{{\it Upper panel}: power of a 10 Mjup planet RV at $P$=10 days added to the observed star RV signal versus the power of that planet alone using the \object{$\beta$ Pic} temporal sampling. {\it Lower panel}: same for a 1 Mjup planet. }
\label{correl_period}
\end{figure}

\begin{figure}
\includegraphics{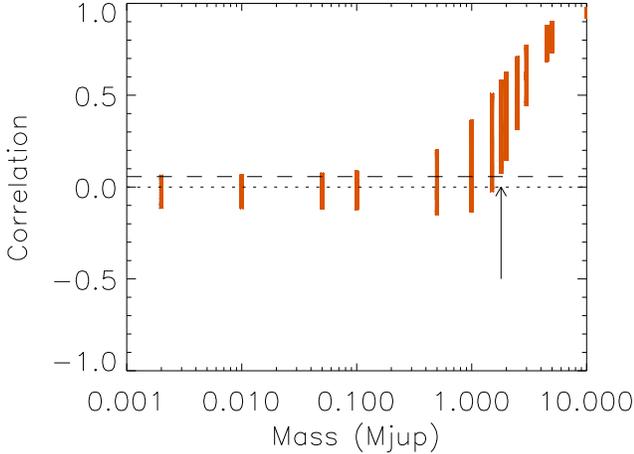}
\caption{Correlation between the power of a planet RV (at $P$=10 days) and the power of that planet added to the star RV, using the \object{$\beta$ Pic} sampling,
illustrating the correlation-based detection limit principle (see text). The dashed horizontal line indicates the threshold, and the arrow the position of the detection limit. }
\label{correl}
\end{figure}

\begin{table*}
\begin{center}
\caption{Detection limit ratios and power criteria }
\label{tabcrit}
\begin{tabular}{cccccccc}
\hline \hline
Star & $RL_{\rm corr}$ & $RL_{\rm peak}$ & $RL_{\rm LPA}$ & $RL_{\rm corrLPA}$ & Domain & $RP_1$ & $RP_2$ \\ 
 & & & & & (days) & & \\ \hline
\object{HD10180}    &  2.79  & 0.88  & 2.32 & 0.43    & 4--60    &   0.64  & 1.93      \\
\object{HD60532}    &  2.14 & 1.30 & 2.14 & 1.00 & 150--650  & -- & 2.61    \\
\object{HD105690}   &  2.48 & 1.42 & 3.48 & 1.40 & 2--5 & 2.53 & 3.29  \\
\object{HD115892}   &  3.23 & 0.76 & 3.28 & 1.06 & 2--5 & 1.56 & 2.10  \\     
\object{HD124850}   &  1.07 & --   & 2.21 & 2.07 & --    & -- & --      \\
\object{HD172555}   &  1.64 & 0.91 & 2.37 & 1.49 & 0.01-0.05 & 1.08 & --    \\
\object{HD199260}   &  2.07 & 0.84 & 1.87 & 1.00 & 1-4  & 0.62 & 1.48  \\
\object{HD210302}   &  1.03 & 1.06 & 1.49 & 1.50 & 1-3  & 0.93 & 1.14  \\
\object{HD219482}   &  1.87 & 0.88 & 2.07 & 1.25 & 1-3  & 1.39 & 2.02  \\
\object{$\beta$ Pic}  &  4.00 & 3.14 & 4.53 & 1.15 & 0.01-0.05 & 2.28 & -- \\ \hline 
\end{tabular}
\tablefoot{
$RL_{\rm corr}$ is the median of the ratio of the rms-based detection limits to the correlation-based detection limits. 
$RL_{\rm peak}$ is the median of the ratio of the rms-based detection limits to the peak detection limits. 
$RL_{\rm LPA}$ is the median of the ratio of the rms-based detection limits to the LPA detection limits. 
$RL_{\rm corrLPA}$ is the median of the ratio of the correlation-based detection limits to the LPA detection limits. 
The domain indicates the period range used in Sect.~6 to estimate the power corresponding to the rms RV. 
$RP_1$ and $RP_2$ are the ratio of the power at the origin of the RV variations (computed, respectively, across the 0.01--1000 and 2--1000 day domains) to the average power threshold used in the LPA method. }
\end{center}
\end{table*}

\begin{table*}
\begin{center}
\caption{Percentage of cases for which the difference from the reference detection limits is smaller than a given threshold. }
\label{tabres}
\begin{tabular}{ccccccccc}
\hline \hline
Method & Test & $<$0.2 Mjup & $<$0.5 Mjup & $<$1 Mjup & $<$1\% & $<$5\% & $<$10\% & Nb values \\ \hline
Rms & & & & & & & \\
   & Nb realizations  &  100   &  100   & 100    &  100   &  100   &   100  &  600   \\ \hline
Correlation & & & & & & & \\
  & Nb realizations & 92.8  & 97.1 & 98.6 & 74.3 & 90.0 & 95.7 & 70 \\
  & Nb freq 8000 & 99.7  & 100.0 & 100.0 & 96.7 & 99.7 & 100.0 & 600\\
  & Nb freq 5000 & 98.7  & 99.7  & 99.8  & 95.7 & 99.5 & 99.8  & 600\\
  & Nb freq 3000 & 95.8  & 99.5  & 99.8  & 91.8 & 98.5 & 99.2  & 600\\
  & 1--500 days & 86.4  & 96.6 & 99.1    & 60.0 & 78.4 & 89.3 & 560 \\
  & 2--500 days & 97.0  & 97.7 & 97.9    & 86.1 & 93.6 & 94.8 & 560 \\
  & 1--100 days & 89.7  & 97.2 & 99.7    & 59.7 & 78.2 & 87.2 & 320 \\
  & 2--200 days & 96.7  & 98.9 & 100.0   & 77.9 & 90.2 & 92.6 & 460 \\ \hline
Peak & & & & & & & \\
  & Nb realizations & 95.1  & 95.1 & 95.1 & 92.7 & 95.1 & 90.2 & 41 \\
  & 1 peak                 & 75.9  & 90.7 & 91.7 & 35.1 & 46.2 & 55.1 & 316\\
  & 3 peaks                & 84.8   & 92.2 & 92.8 & 60.9 & 72.2 & 84.9 & 345\\
  & 7 peaks                 & 85.7  & 92.7 & 93.1 & 66.7 & 80.6 & 88.0 & 361 \\
  & Exclusion period window & 99.4  & 99.7 & 99.7 & 98.9 & 99.4 & 99.7 & 350 \\
  & Planet period window   & 93.6  & 95.3 & 97.5 & 89.2 & 90.6 & 92.5 & 361 \\ \hline
LPA & & & & & & & \\ 
  & Nb realizations & 100  & 100 & 100 & 98.6 & 100 & 100 & 70  \\
  & Nb freq 8000    & 94.5  & 98.5 & 99.7 & 80.3 & 96.8 & 98.5 & 600  \\
  & Nb freq 5000    & 91.5  & 96.7 & 98.3 & 76.0 & 90.2 & 95.7 & 600  \\
  & Nb freq 3000    & 83.2  & 91.5 & 95.2 & 56.3 & 73.0 & 85.8 & 600  \\
  & 1--500 days     & 91.9  & 96.4 & 98.9 & 79.1 & 91.6 & 95.7 & 560  \\
  & 2--500 days     & 98.0  & 100 & 100 & 88.2 & 98.6 & 99.3 & 560  \\
  & 1--100 days     & 94.7  & 97.8 & 99.1 & 75.0 & 90.3 & 94.7 & 320  \\
  & 2--200 days     & 96.3  & 99.6 & 99.8 & 86.9 & 97.4 & 98.5 & 464  \\ 
  & Window          & 80.2 &  90.7 & 95.3 & 54.3 & 62.3 & 73.0 & 600  \\ \hline
\hline
\end{tabular}
\tablefoot{The reference detection limit corresponds to the parameters used at the begining of each section (respectively Sect.~3.1, 4.1, 5.1, and 6.1).}
\end{center}
\end{table*}

The correlation-based method takes into account that the presence of a planet at a period $P$ induces significant power at the planet frequency but also 
at other frequencies,
depending on the time sampling and the planet period $P$. 

Fig.~\ref{period} shows the periodogram of a RV signal $RV_{\rm pla}$ 
for a 10 Mjup planet alone at $P$=10 days using
the \object{$\beta$ Pic} temporal sampling (upper panel), for periods between 2 and 1000 days\footnote{By default, the number of frequencies used to compute 
the periodogram is 10000, and the periodograms are computed for periods between 2 days and 1000 days, which were chosen to enclose all the considered periods}:  
the planet peak is much stronger than the stellar signal, as well as the power at all periods. This planet is massive, and when its corresponding RV is added to the stellar RV signal $RV_{\rm star}$, 
the periodogram (Fig.~\ref{period}, middle panel) is then very similar to the first one. However, the periodogram for a less massive planet (i.e. 1 Mjup for example, which we find to be lower than the detection limit) is
very different, as shown in the lower panel of Fig.~\ref{period}, and is of course dominated by the stellar signal. 

The correlation between the periodograms computed for the planet alone $RV_{\rm pla}$ 
and the planet added to the stellar signal $RV_{\rm pla}$+$RV_{\rm star}$ is therefore close to 1 when the planet dominates the signal (planet above the detection limit), as shown
in the upper panel of Fig.~\ref{correl_period}, while it is close to 0
when the stellar signal dominates the signal (planet below the detection limit), as shown in the lower panel of Fig.~\ref{correl_period}. 
We then infer the detection limit
from the variations in this correlation as a function of the planet mass. This function is illustrated in Fig.~\ref{correl} for $P$=10 days, for 100 realizations at each planet mass
(corresponding to various phases of the planet RV signal). 
The threshold is computed as the maximum of the correlations obtained for a very low planet mass (0.002 Mjup here).

We then define the detection limit as the minimum mass for which correlation values
are above this threshold for all realizations (spanning phases between 0 and 2$\pi$). With 100 realizations, this detection limit is therefore at a confidence level of 1/100 (99\%). 
This method is time-consuming, hence the choice of the 1/100 confidence level instead of 1/1000. In the next section, we compare our results
with 1000 realizations. The mass step is about 0.1 Mjup (about 0.05~Mjup in the case of \object{HD105690}) and 0.01 Mjup for \object{HD10180}.
The impact of the period range used to compute the periodograms is studied in Sect.~4.2.2.

\begin{figure}
\includegraphics{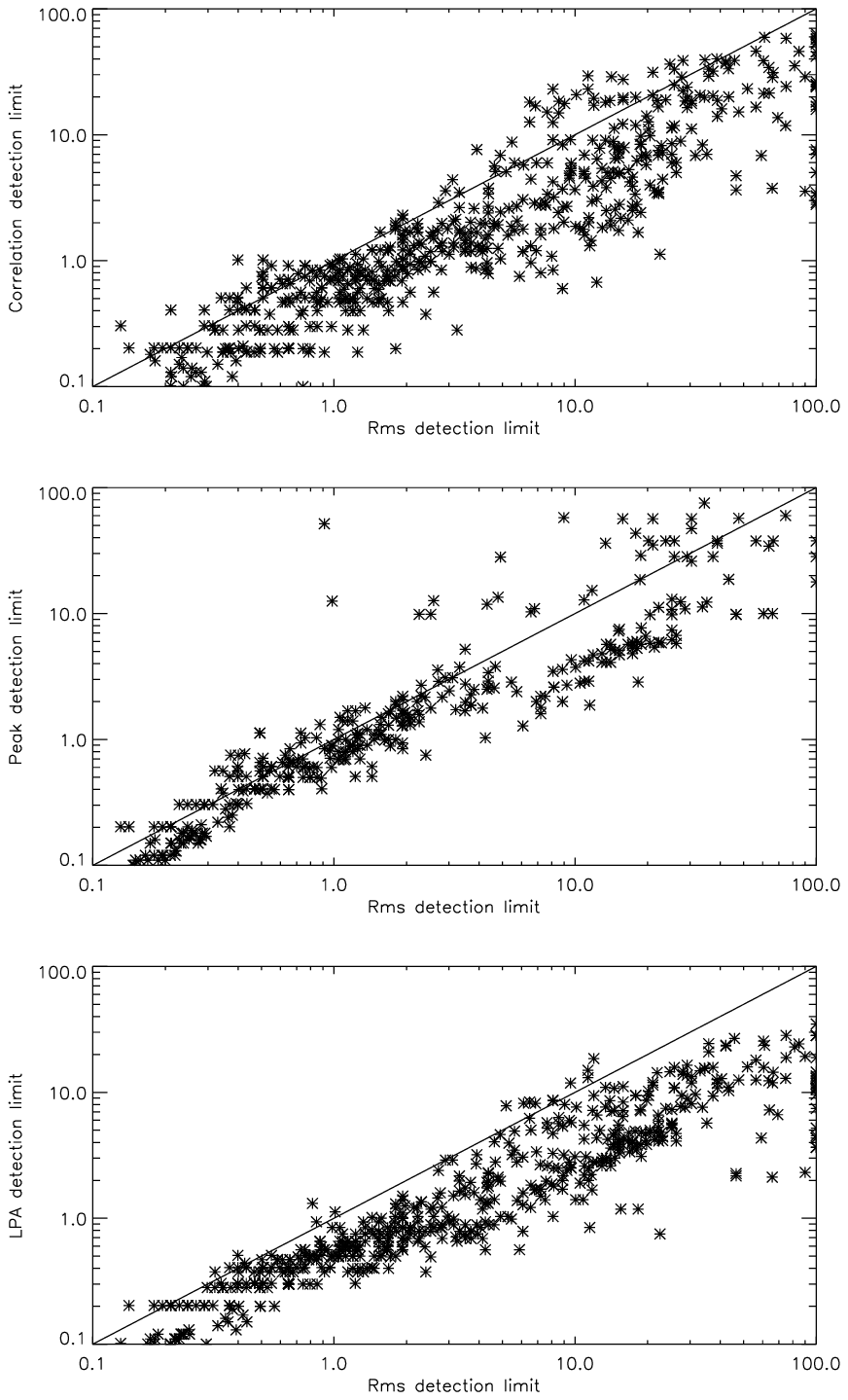}
\caption{{\it Upper panel}: correlation-based detection limit versus rms detection limit (in Mjup) for all stars and periods. The solid line indicates the 
y=x line. 
{\it Middle panel}: same for the peak detection limits. 
{\it Lower panel}: same for the periodogram detection limits.
}
\label{crit4}
\end{figure}

\begin{figure}
\includegraphics{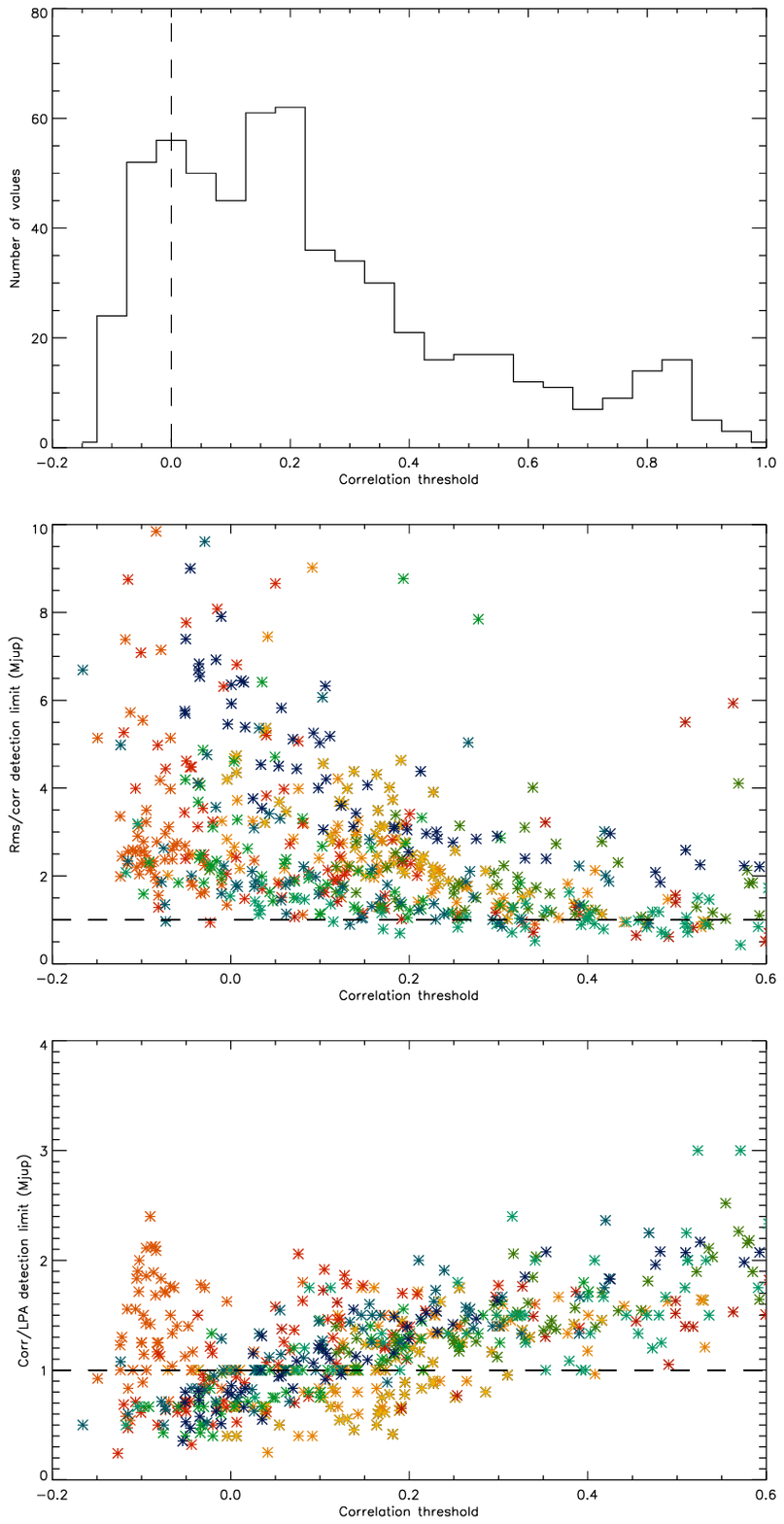}
\caption{
{\it Upper panel}: distribution of the correlation threshold for all stars of the sample and periods. 
{\it Middle panel}: ratio of the rms detection limit to the correlation-based detection limit for all stars (one color per star) and periods, versus the correlation threshold.
{\it Lower panel}: ratio of the correlation-based detection limit to the LPA detection limit for all stars (one color per star) and periods, versus the correlation threshold.} 
\label{seuil}
\end{figure}

The resulting detection limits for the ten stars  are shown in Fig.~\ref{res_a} (right panels, orange stars). They are compared to the rms-based method, 
as well as the other methods described in more detail in Sect.~5 and 6. In this section, we focus on the comparison with the rms-based method.
Fig.~\ref{crit4} compares the correlation-based detection limits with
the rms detection limit for all stars and periods. The correlation-based method provides detection limits that are lower than the rms method in most cases. 
The ratio of the two is typically between 1 and 10 (the correlation-based method therefore leading to a possible improvement of up to one order of magnitude).  

To quantify the difference in detection limit between the methods, we compute for each star the median (over periods) ratio of the rms detection limit to 
the correlation-based detection limit, $RL_{\rm corr}$. 
A small $RL_{\rm corr}$  (close to 1) means that both methods give similar detection limits, while a large $RL_{\rm corr}$ means that 
the correlation-based method significantly improves the detection limits. The values of  $RL_{\rm corr}$ are shown in Table~\ref{tabcrit}. 
The best improvement on the rms method is obtained for \object{$\beta$ Pic}, which can be explained by the strong pulsation signal (at the origin of the observed 
rms RV) being significantly larger than the rest of the power in the periodogram. This is followed by the results \object{HD115892}, for  which there is 
a strong improvement as well. 
We expect the correlation-based method to give lower estimates of the detection limits than the rms method when most of the RV dispersion gives power 
in a period range outside the periods that we considered. This is why \object{HD172555} has a relatively low value of $RL_{\rm corr}$: the power due to the 
pulsations is no greater than the power at longer periods. To quantify this relationship, it is necessary to derive a criterion based on the ratio of the 
power at the origin of the observed rms RV to the power in the periodogram: this is done in more detail 
in Sect.~7. 

Finally, the correlation-based method is very sensitive to the threshold applied to the correlation. We recall that the correlation threshold is taken as 
the strongest correlation between the planet RV periodogram and star + planet RV periodogram for a very low mass (and 100 realizations of the phase). 
In many cases, the correlation threshold is close to zero, as shown in Fig.~\ref{seuil} (upper panel). However, in some cases, the threshold is large: for 
a particular phase, the planet periodogram happens to be correlated with the observed signal. This is likely to depend strongly  on the temporal sampling 
of the observations. Fig.~\ref{seuil} (middle panel) shows the ratio of the rms to correlation-based detection limits versus the correction threshold: 
small thresholds are indeed associated with the best improvement, while the ratio tends towards 1 (no improvement) when the threshold is large. The correlation-based method is therefore efficient mostly when the threshold is small.

\subsection{Impact of the parameters}

\subsubsection{Number of realizations}

\begin{figure}
\includegraphics{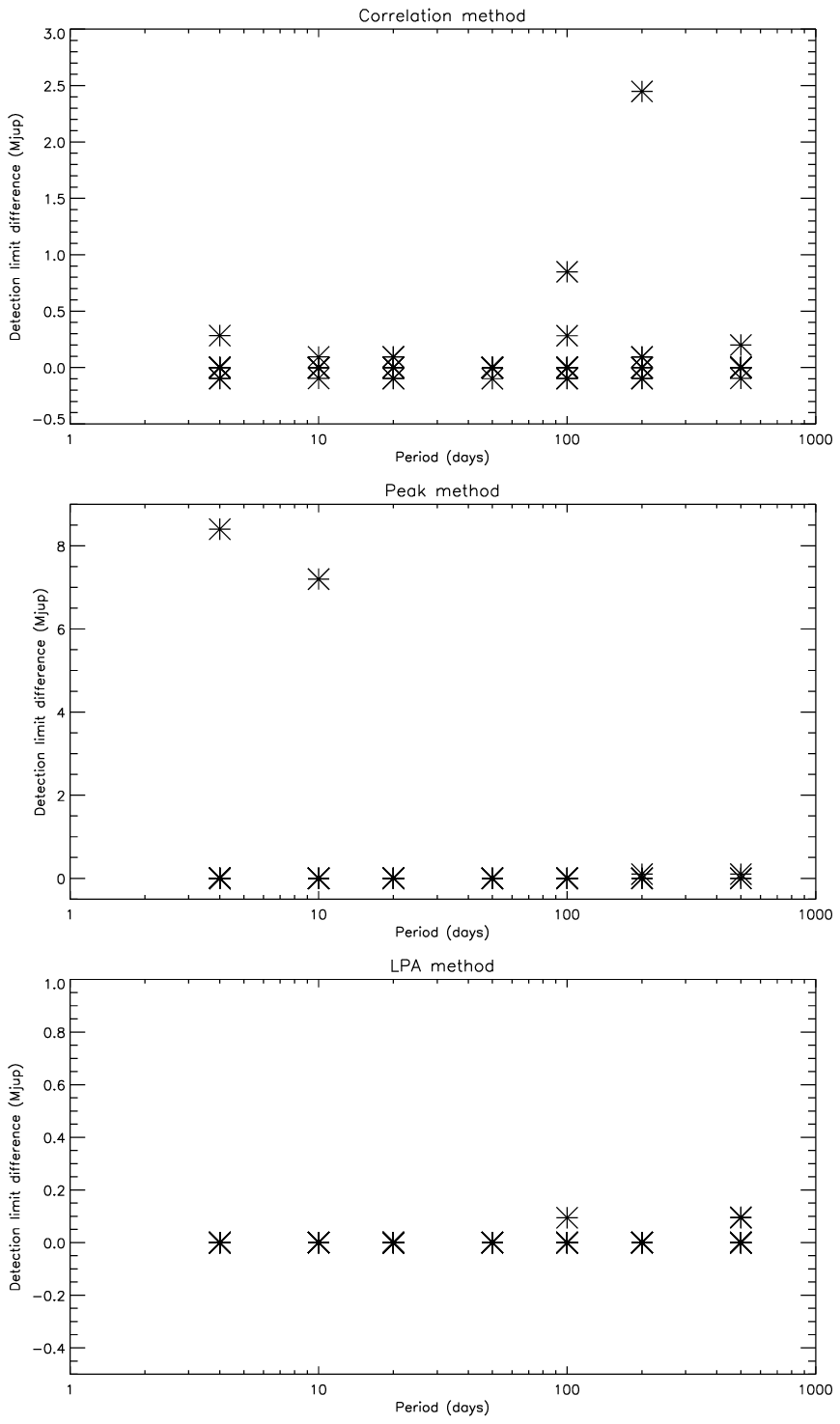}
\caption{ {\it Upper panel}: difference of detection limit between 1000 and 100 realizations, for the correlation-based method, all stars, 
for a selected number of periods.
{\it Middle panel}: same for the peak method.
{\it Lower panel}: same for the LPA method.}
\label{nbreal}
\end{figure}

As pointed above, this method is time-consuming because it involves the computation of many periodograms. Our algorithm uses variable mass steps of decreasing
values in order to converge faster, but it remains slow and of course depends strongly on the number of realizations. A value of 1000 realizations would be necessary
to derive a detection limit with a 99.9\% confidence. Here we therefore compare the detection limits computed using 100 and 1000 realizations for a subset of 7 periods
(4, 10, 20, 50, 100, 200, and 500 days) spanning the whole range of periods. The difference between the two detection limits is shown in Fig.~\ref{nbreal} for the ten stars. Table~\ref{tabres} gives the percentage of points that differ by more than a certain value (either absolute or relative) from the reference detection limits obtained in Sect.~4.1. 
Two-thirds of the detection limits are strictly identical for the two computations. Eight percent of the points differ by more than the 0.1 Mjup mass step, the two largest (difference of 5--6\%) corresponding to periods of 100 days and 200 days, respectively, for \object{HD124850} and \object{HD172555}. 
These differences are smaller than the dispersion observed in the plot of the detection limit versus period for these stars. 
We therefore conclude that the use of 100 realizations instead of 1000, which represents a considerable gain in computing time, is justified, leading to uncertainties smaller than a few percent, and most of the time much better (i.e. on the order of our mass step).

\subsubsection{Periodogram parameters}

\begin{figure}
\includegraphics{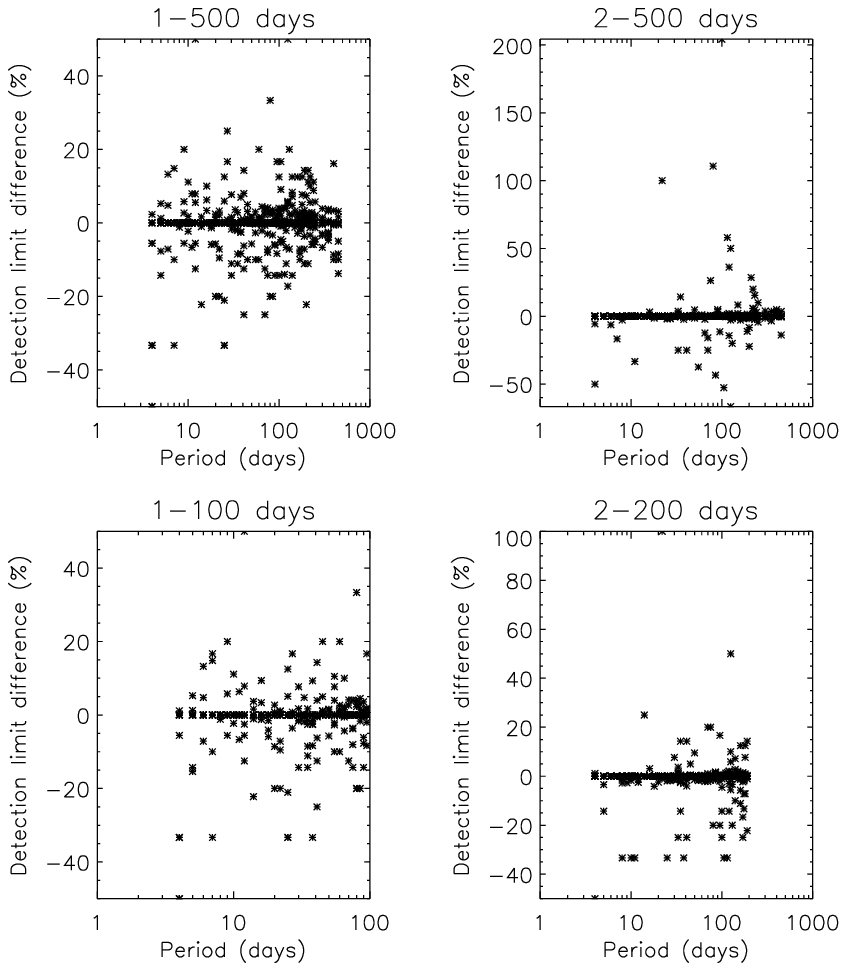}
\caption{Relative difference between the detection limit computed for the period range indicated compared to the reference detection limit defined in Sect.~4.1.1 for the correlation-based method,
where we plot data for all stars.}
\label{pminmax}
\end{figure}

The other parameters used in the correlation-based method are the number of frequency elements used to compute the periodogram, and the period range considered.
We studied their impact on the results separately, by computing the detection limit for the 60 periods and for each star with different numbers of frequency elements or different period domains. 
The results of the tests are shown in Table~\ref{tabres}.

For the reference period domain (i.e. 2--1000 days), we computed all detection limits for three numbers of frequency elements: 3000, 5000, and 8000. We found that
the difference from the previous detection limits (10000 frequency elements) is smaller than 0.5~Mjup in $\sim$99\% of the cases. For 8000, 5000, and 3000 frequency elements respectively,
99.7\%, 99\%, and 96\% of the cases differ by less than 0.2~Mjup.
We therefore conclude that our results do not depend strongly on the choice of frequency number. It may be possible to reduce the computation time by 
using a smaller number of frequency elements without significantly affecting the results. 

Finally, we considered the impact of the period range on the detection limits. We tested four other period ranges: 1--500 days, 2--500 days, 1--100 days, and 2--200 days. 
The detection limits are computed only for the periods included in the corresponding period range. The results are shown
in Fig.~\ref{pminmax}. When all period ranges are considered, the difference observed is below 0.5~Mjup in more than 96\% of the cases. 
The percentages of values exhibiting differences smaller than 0.2 Mjup are still large,
with percentages of $\sim$97\% and $\sim$96\% for 2--200 and 2--500 respectively, and $\sim$86\% and $\sim$89\% for 1--500 and 1--100. The difference from the reference detection limits 
is therefore slightly larger when including the 1--2 day period range. This is due to a strong peak around one day in all time series, owing to the observing pattern. This pattern being always present, 
it may be more efficient not to include this period domain in the computation of the periodogram as it does not help us to discriminate between various patterns.  
Here again we conclude that the choice of period parameters does not strongly affect our results and gives an idea of the uncertainty in the detection limit, although it may be preferable to eliminate period domains exhibiting a pattern in the periodograms present in all configurations.

\subsection{Conclusion about the correlation-based method}

The correlation-based method provides detection limits that are lower than the rms method in most cases. 
The choice of the number of frequency elements has a very small impact on the resulting detection limits. 
That of the period domain has a larger impact, but mostly when including the one day period: this domain has significant power owing to the temporal sampling and therefore it should be easier to discriminate the presence of the planet while not considering this period domain. 
Since minimizing the amount of computing time is critical, we allow the use of the 1/100 confidence level, which allows us to retrieve a very precise 
estimation (with an uncertainty smaller a few percent) in most cases. We conclude that the uncertainties are on the order a few percent for most cases. 
 

\section{Peak amplitude method}

\subsection{Method and results}

\begin{figure}
\includegraphics{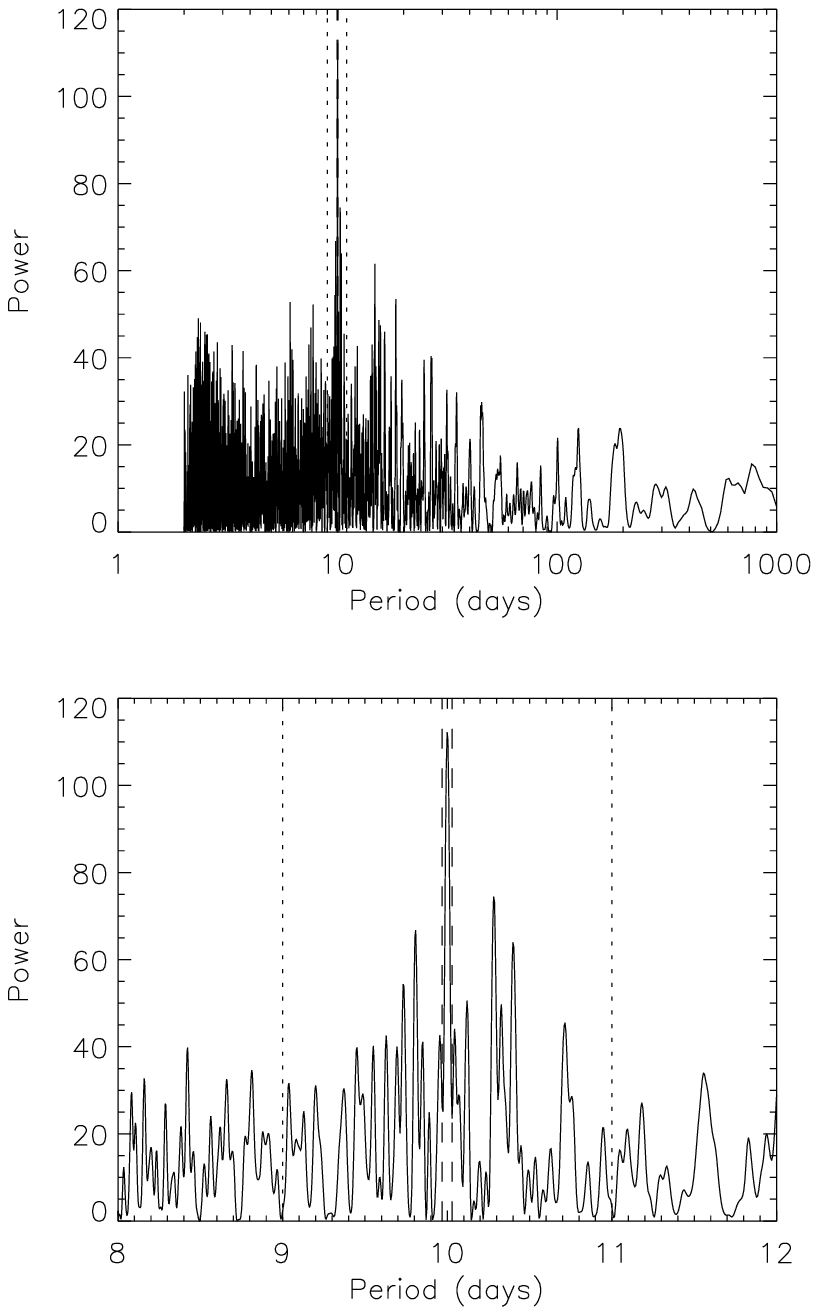}
\caption{ {\it Upper panel}: periodogram of the RV signal of a 10 Mjup planet of period $P$=10 days for the \object{$\beta$ Pic} sampling. The dotted vertical lines shows the position of
the exclusion period window (see text) and the dashed vertical lines the position of the planet period window (see text) used in the peak method. {\it Lower panel}: zoom for the period range 8--12 days.}
\label{period_pic}
\end{figure}

\begin{figure}
\includegraphics{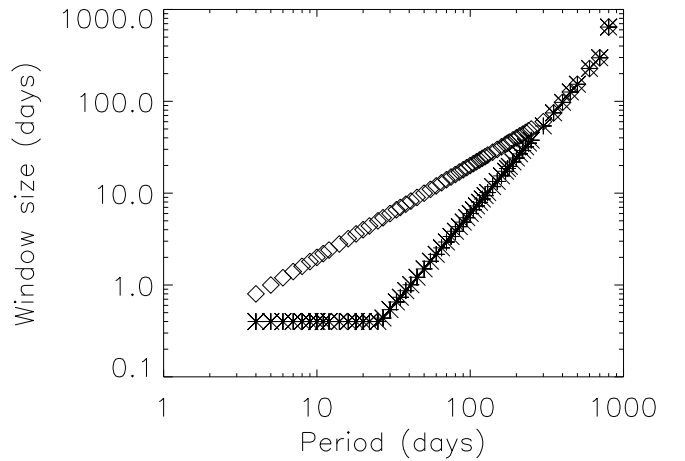}
\caption{Planet-period window (stars) and exclusion-period window (diamonds) versus the period for the peak method. }
\label{win}
\end{figure}

\begin{figure}
\includegraphics{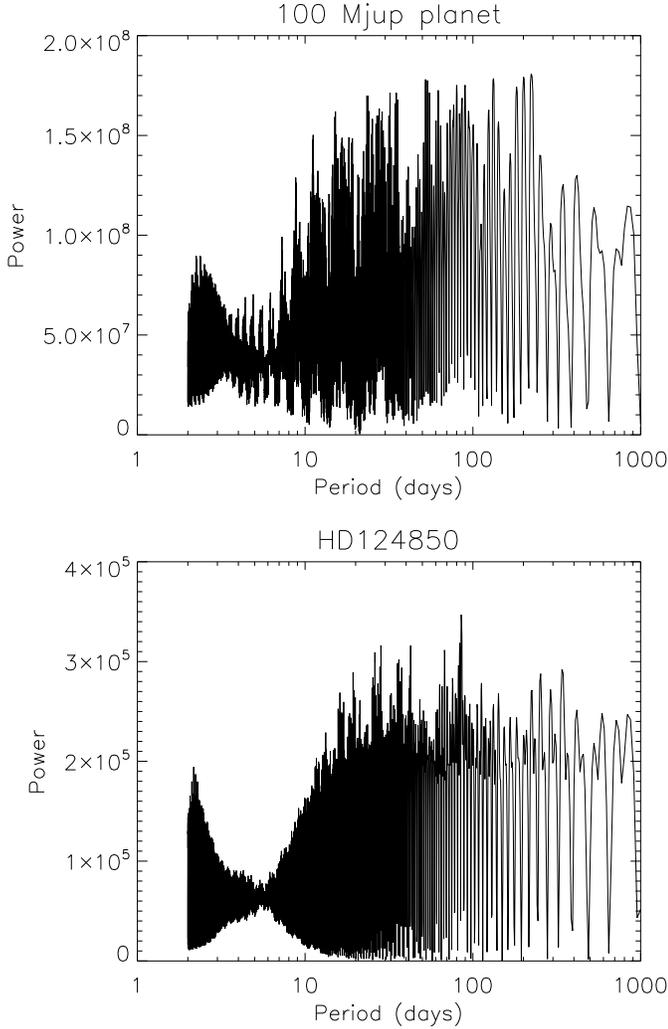}
\caption{ {\it Upper panel}: periodogram for a 100 Mjup planet of period $P$=100 days with the time sampling of \object{HD124850}. {\it Lower panel}: periodogram for the observed star RV. }
\label{per_plan}
\end{figure}

\begin{figure}
\includegraphics{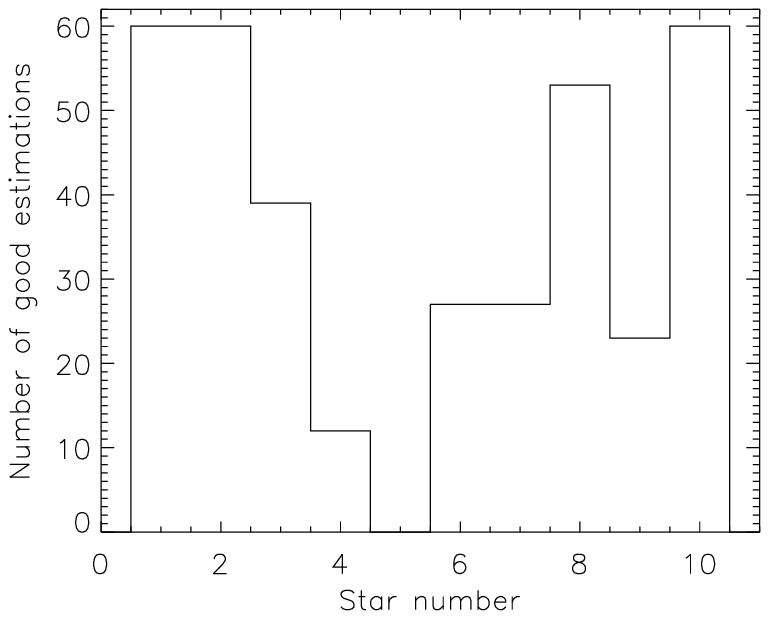}
\caption{Number of valid peak detection limits versus the star number (see Table~1, first column).}
\label{nbpicok}
\end{figure}

The principle of this method is to compare the amplitude of the peak at the planet period with the average of the amplitude of
the five next highest peaks that are not at the planet period.

This method is close to the procedure usually used to actually detect a planet on a observed signal. We add a planet RV (with a given mass, period, and phase) 
to the observed star RV and compute the periodogram. We first identify the peak corresponding to the planet and measure the amplitude of this peak. We then 
identify the five largest peaks that are not at the planet period, and compute the average amplitude. For a given period and mass, 100 realizations of the 
phase are produced. We search for the lowest mass such that the planet peak amplitude is larger than other peaks for all realizations. 
Planet identification is usually based on this approach. 
However, it requests several parameters before it can be used automatically. In this paper, we use the following parameters:

\begin{itemize}
\item{The periodogram parameters including the range of periods over which the periodograms are computed and therefore the peaks identified, and the number
of frequency elements in the periodogram. We take 2--1000 days, as for the correlation-based method, and 10000 as the number
of frequency elements.}
\item{The range of periods  over which the planet peak is searched for (hereafter the planet period window). As the temporal sampling is imperfect, the period at which the
peak corresponding to the planet period is found (even for a massive planet) is usually not exactly the input period. Half the period window size is taken as the maximum value between 
6 frequency resolution elements\footnote{We determined this amplitude empirically by measuring the position of massive planet peaks for all stars in the sample.} and 0.2 days\footnote{This minimum is also found empirically by checking the size of the window at very small planet periods.}. Fig.~\ref{period_pic} shows an example of the planet period window. The size of the planet period window as a function of planet period is shown in Fig.~\ref{win}.}
\item{The period range outside of which the 5 highest peaks are searched for (hereafter the exclusion period window). This period range must not include peaks that are too close
to the input planet period. We define the lower and higher limits as 0.9 and 1.1 times the input period of the planet. This choice is driven by 
there usually being many large peaks around the peak closest to the true planet period.
The exclusion period window is increased to match the planet period window when the planet period window happens to be the largest of the two. The full size 
of the exclusion period window is shown in Fig.~\ref{win}. We note that this is a simplified definition compared to 
that used in paper I (see Sect. 5.2.3). Fig.~\ref{period_pic} shows an example of the exclusion period window. }
\item{The number of realizations, which related to the confidence level of the detection limit (our default value is one out of 100). This method is time-consuming and therefore it is necessary to optimize the number of realizations. For a given mass, 100 phases for the planet are considered.  }
\item{The number of peaks over which we average the peak amplitudes (i.e. for peaks outside the planet period window). In our standard method, we consider five peaks. This choice is a good compromise to limit the uncertainty on the detection limit. The impact of this parameter is studied in Sect.~5.2.2. }
\end{itemize}

The resulting detection limits are shown in the right panels of Fig.~\ref{res_a} for the ten stars (red diamonds). 
This method works well only when the data are homogeneously sampled: 
in some cases (i.e. for some periods and some stars), the method does not converge. This happens when the planet peak is never the largest one, even for a 
very massive planet, owing to the temporal sampling. An illustration of this is shown in Fig.~\ref{per_plan} for \object{HD124850}. The RV variations
for this star are poorly sampled, as shown in Fig.~\ref{res_a}. Fig.~\ref{per_plan} shows the periodogram for a 100 Mjup "planet" (which is very massive) 
at a period of $P$=100 days: no planet peak is visible owing to the very poor temporal sampling.
The power is however much higher than for the observed RV, demonstrating that a 100 Mjup cannot be present, otherwise the observed power would be much higher. 
This is therefore a first limitation of the peak method. We do not consider these cases in the following analysis and test the robustness of the method for the remaining points, which are the only ones shown in Fig.~\ref{res_a}. 

Fig.~\ref{nbpicok} shows the number of relevant periods (out of 60) for each star. \object{HD124850} has no suitable detection limits, while \object{HD115892} has very few.
\object{HD172555}, \object{HD199260} and \object{HD219482} have detection limits for only about half the periods. This means that depending on the temporal sampling, a planet, 
even a massive one, can have a significant impact on both the signal and the periodogram without producing a peak, which justifies the use of the other methods.
Since this method is often used to detect planets, one should be aware that some detections may be missed if one relies only on peak amplitudes.

Fig.~\ref{crit4} shows the peak detection limits versus the rms detection limit for all stars and periods. There is usually an improvement on the rms method, 
but there are a significant number of points for which the peak detection limit is worse. As for the correlation-based method, we computed a ratio 
$RL_{\rm peak}$ to quantify this improvement. The results are shown in Table~\ref{tabcrit}. Half of the stars show an improvement, while the other half 
have a higher detection limit. The greatest improvement is by far achieved for \object{$\beta$ Pic}.

\subsection{Impact of the parameters}

\subsubsection{Number of realizations}

As in Sect.~4.1.2 for the correlation-based method, we compare the detection limits computed using 100 and 1000 realizations for a subset of seven periods
(4, 10, 20, 50, 100, 200, and 500 days) spanning the whole range of periods. The differences between the two detection limits for all stars are shown 
in Fig.~\ref{nbreal}. A summary of the test is shown in Table~\ref{tabres}.  
Only two points, representing 3\% of the values, indicate that there is any significant difference and are certainly related  
to the absence of a planet peak even for quite massive planets for certain samplings and a significant number of phases.
We conclude that the use of 100 realizations instead of 1000, which represents a considerable gain in computing time, is justified.

\subsubsection{Number of peaks outside the exclusion period window}

\begin{figure}
\includegraphics{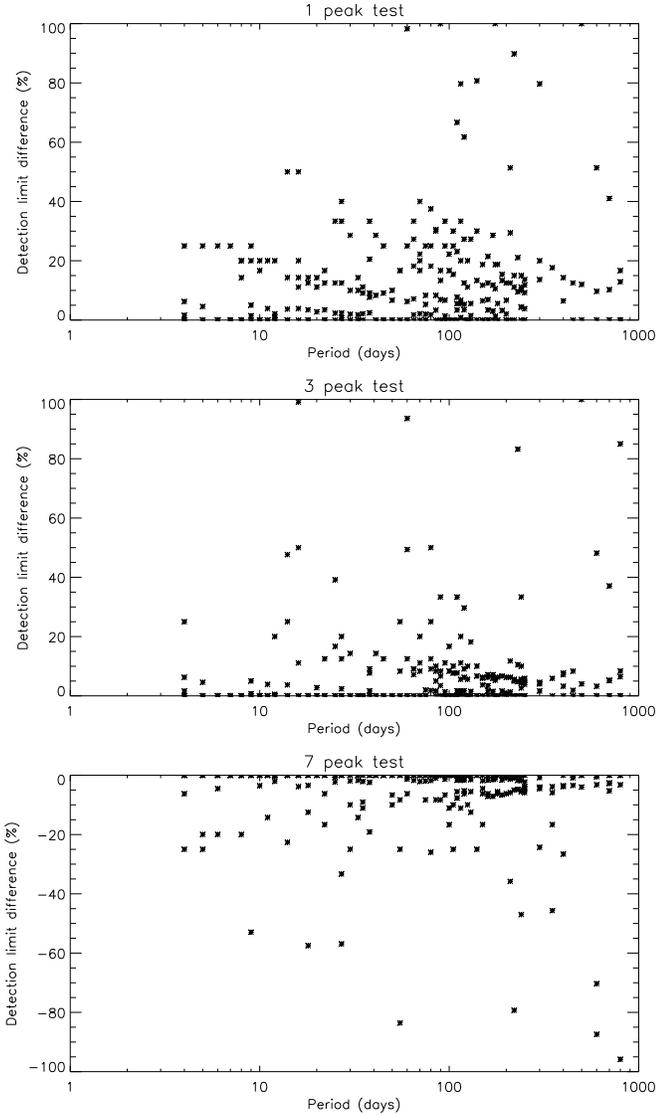}
\caption{{\it Upper panel}: relative difference between the one peak computation and the reference detection limit (i.e. five peaks, with other parameters 
described in Sect.~5.1). A few points are off the 100\% scale for clarity.
{\it Middle panel}: same for three peaks instead of one peak. {\it Lower panel}: same for seven peaks instead of one peak. }
\label{nbpic3}
\end{figure}

We averaged the amplitudes of the five highest peaks outside the exclusion period window to limit the noise in the result. We now study the impact of the number of 
peaks on the resulting detection limits. We therefore compute the detection limit for nine stars (all except \object{HD124850}) as a function of the period, for one 
(i.e. no averaging), three, and seven peaks. The results are shown in Table~\ref{tabres}. There is good agreement between the different time series. 
Fig.~\ref{nbpic3} compares the differences to the five-peak computation for all stars. As expected, the detection limit is higher for one peak and three 
peaks compared to the five-peak computation, while it is lower for the seven-peak computation. 
The differences are slightly larger for \object{HD111998}, \object{HD172555}, and \object{$\beta$ Pic} at long periods. 
For three and seven peaks, respectively, $\sim$92\% and $\sim$93\% of the points
have a difference smaller than 0.5~Mjup, showing that the detection limit is quite insensitive to the number of peaks across which we average. 
The percentage is 91\% for one peak so the difference is slightly larger in  that case.

\subsubsection{Size of the exclusion period window}

In paper I, we used a more complex method to determine the size of the exclusion period window than in this paper. Instead of using a simple percentage of the
period, we determined an intensity threshold in the periodogram and used this threshold to determine the exclusion period window, with a maximum set at $\pm$10\%
from the planet period. We recomputed all detection limits using that rule instead of the simple 10\% rule used above.
The results are shown in Table~\ref{tabres}.
We find that for more than  99\% of the values the difference is below 0.2 Mjup. The agreement is therefore very good, which justifies the use 
of the simplest and more robust method.

\subsubsection{Size of the planet period window}

Finally, we check the impact of the size of the planet period window on the results by recomputing the detection limit using a window twice as large. 
The results are shown in Table~\ref{tabres}.
We find that for almost 94\% of the values, the difference is smaller than 0.2 Mjup. 
We note that increasing the planet period window naturally increases the number of valid points, but only by a small amount.

\subsection{Conclusion on the peak method}

The peak method usually provide some improvement on the rms method, but for only half of the stars on average in the sample. 
The most important limitation of this method is that for some temporal samplings, the amplitude of a planet peak, even for massive planets, is no larger 
than the other peaks. For a given star, this was shown to happen for only some periods, or even for all periods, as seen for one star. Varying the parameters does not significantly impact this result. 
In the remaining cases, the impact of most parameters is small, as shown for the number of realizations, and sizes of both the exclusion and planet period windows. The most sensitive parameter is the 
number of peaks chosen to compare with the planet peak. The method is quite insensitive to the number of peaks over which we average, i.e. above three peaks, 
but if no averaging (one peak only) is used the detection limits are significantly increased.
The uncertainties are larger than for the correlation-based method, and we can estimate them to be on the order of 10\% in most cases.

\section{LPA method}

\subsection{Method and results}

\begin{figure}
\includegraphics{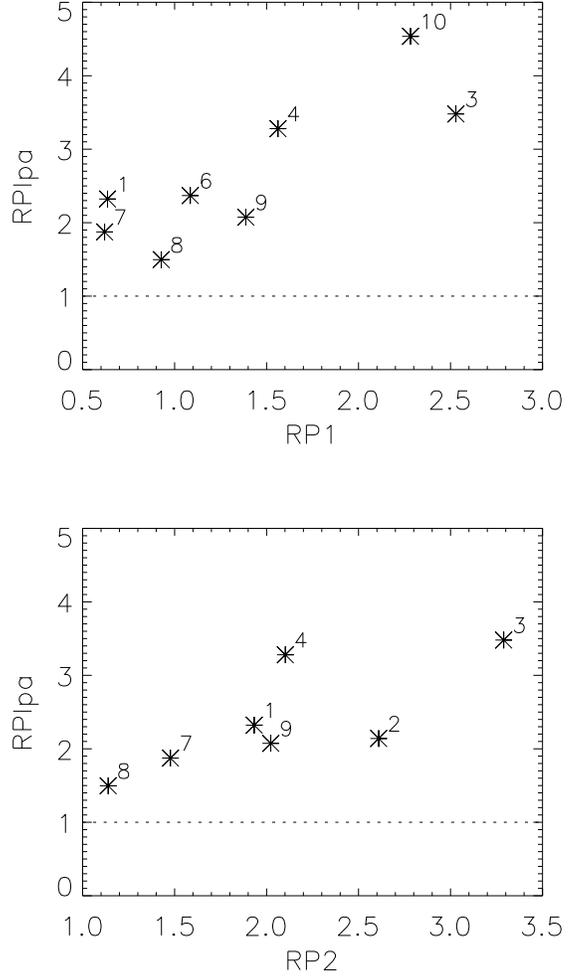}
\caption{{\it Upper panel}: median ratio of the rms detection limit to the LPA  detection limit $RL_{\rm LPA}$ versus the power ratio $RP_1$ (see text, 
one point per star). The number adjacent to each cross gives the star number as listed in Table~1. 
{\it Middle panel}: same versus the power ratio $RP_2$ (see text). 
}
\label{crit}
\end{figure}

The rms method has the drawback that the rms of a given RV signal includes contributions from all frequencies, although most of this power
may come from frequencies far from the planet period investigated. On the other hand, the peak method described in the previous section, which is close 
to the method used to 
detect planets in an observed signal, is less robust than the other methods presented here and relies on many parameters. 
A strong limitation of the rms method is that, in some cases, if the sampling is inadequate, even a high mass
planet may not produce any significant peak. However, a massive planet does produce a strong power in the periodogram. We present here a method that is based on  
the best of these two methods: 
the principle is, for a given planet and period, to compare the power of the signal produced by this planet alone with the power of the actual signal within a localized period range. 
The basic idea is that a detection requires the planet-induced power to be higher than the power of the actual signal. In Fig.~\ref{crit4}, we could for 
example see that despite the planet peak being unable to be identified, the power due to the planet RV is much higher than the observed power, meaning that 
the presence of a 100 Mjup companion could clearly be ruled out.
Hereafter, we call this method the LPA ({\it local power amplitude}) method.

The principle of the LPA method is the following: to compute
the detection limit for a given planet period $P$, we first measure the maximum power in the observed periodogram around this period, e.g. in the range 
$[$0.75$P$--1.25$P]$\footnote{The range is chosen to be small enough to study a specific range of periods, and large enough to enable us to
compare the  peak with the closest ones: if the range is too small, the power in that domain may be much lower by chance than very nearby peaks, leading to an 
underestimation of the detection limit, and the planet peak cannot be compared with its environment.}, which is hereafter denoted as $Pow_{\rm obs}$. 
For a planet with a period $P$ and a given mass $M$, we compute the RV signal for the same temporal sampling, then its
periodogram in the period range 2--1000 days over 10000 frequency elements as before, and then compute the maximum
power in the same range, $Pow_{\rm pla}$. If for all realizations (100 phases, as for the previous methods), $Pow_{\rm pla}$ is above $Pow_{\rm obs}$, 
then we consider this mass $M$ to be above the detection limit. Computation are performed at the 1/100 level, and the impacts of the various parameters are 
tested in the next few sections.

The resulting detection limits are shown in Fig.~\ref{res_a}. As shown in Fig.~\ref{crit4}, there is an improvement on the rms method in most cases. 
As before, we define a criteria to quantify this improvement on the rms method, $RL_{\rm LPA}$, shown in Table~\ref{tabcrit}. 
Most of the time, these detection limits are on average the lowest for all stars.
The ratio also changes in some cases with the periods. For example, for \object{HD60532}, the LPA detection limits is lower (than for the rms method) at small 
periods, but the two curves are close to each other at long periods owing to the presence of the planet peaks. On the other hand, for \object{HD115892}, the 
improvement increases as the period increases, because the power in the periodogram drops significantly from short to long periods.

\subsection{Impact of the parameters}

\subsubsection{Number of realizations}

As in Sect.~4.1.2 for the correlation-based method, we compare the detection limits computed using 100 and 1000 realizations for a subset of seven periods
(4, 10, 20, 50, 100, 200, and 500 days), spanning the whole range of periods. The results are shown in Table~\ref{tabres}. The differences between the two 
detection limits is equal to zero for almost all points, except for three points where the difference is about 0.1 Mjup (i.e. our mass step). The agreement, 
shown in Fig.~\ref{nbreal}, is therefore excellent and shows that for this method the use of 100 realizations instead of 1000, which represents a 
considerable gain in computing time, is again fully justified.

\subsubsection{Periodogram parameters}

\begin{figure}
\includegraphics{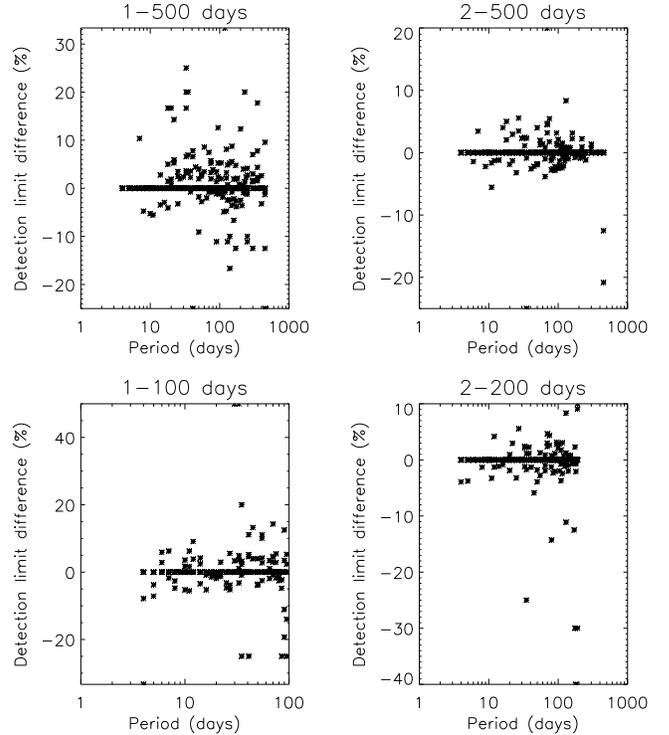}
\caption{Differences between the detection limits computed for the various period ranges indicated compared to the reference detection limit defined in 
Sect.~4.1.1 for the correlation-based method, where we present data for all stars.}
\label{pminmax2}
\end{figure}

For the reference period domain (i.e. 2--1000 days), we compute all detection limits for three numbers of frequency elements: 3000, 5000, and 8000. The 
results are shown in Table~\ref{tabres}. For 8000, 5000, and 3000 frequency elements, respectively, $\sim$98.5\%, $\sim$97\%, and $\sim$91.5\% of all 
detection limits (computed for all stars and periods) differ from the reference detection limits defined in Sect.~6.1 by less than 0.5~Mjup. This method is therefore slightly more sensitive to the number of frequency elements than the correlation-based method, but the agreement remains very good. 

As for the correlation, we also compute the periodograms for different period ranges. The results are shown in Table~\ref{tabres} and Fig.~\ref{pminmax2}. Overall, more than $\sim$96\%  of the points 
differ by less than 0.5~Mjup. About 90\% or more have differences smaller than 5\%. As for the correlation-based method, the worse cases happen when including the 1--2 day range, but the agreement remains very good.

\subsubsection{Period range around planet period}

We used the period domain 0.75$P$--1.25$P$ (where $P$ is the planet period) to compute the maximum power. Here we investigate the influence of this choice 
on the result and test a range that is twice as large, i.e. periods covering the range 0.5$P$--1.5$P$. About 91\% of the values display differences smaller 
than 0.5 Mjup, while 73\% have differences smaller than 10\%.  The result is therefore sensitive to the period range, although not dramatically. 
For example, if in a given period range the observed periodograms have high powers, we know that this will influence the computation of the detection limits 
for periods of up to half the window below or above that period range. 

\subsection{Conclusions about the LPA method}

The LPA method provides the most significant improvement on the rms method. 
The LPA method is insensitive to the number of realizations considered. The number of frequency elements and the period range have a small impact on the resulting detection limits as well  and produce an uncertainty that is on the order of a few percent. The size of the window is the least robust parameter and causes large uncertainties, but is also easy 
to analyze as the threshold can be derived directly from the periodogram of the observed RV.

\section{Discussion}

\subsection{A new criteria for interpreting the detection limits}

We now study in more detail the origin of the improvement relative to the rms method. We observe that a strong improvement is not widespread, as for 
example, for  a star such as \object{HD172555}, which has the strongest rms RV (owing to pulsations), there is no strong improvement with the new methods. 
The rms method is based on the rms RV only, regardless of its origin. This rms RV basically corresponds to a certain power in the periodogram depending on the period at which the 
power is injected, which leads to the increasing detection limit with the period $P$ observed in Fig.~\ref{res_a}. However, the periodogram of 
the observed RV includes some power related to this rms RV (which is more or less visible depending on the period range over which the periodogram is 
computed), but also some power due to the temporal sampling. The LPA method (as for the two 
other methods) relies on the power in the periodogram, and is therefore affected by the total power, i.e. {\it both} the contributions of the rms RV 
and the temporal sampling. We therefore expect to see an improvement on the rms method (i.e. a large $RL_{\rm LPA}$ 
in the present case) when the power due to the rms RV is high compared to the total power in the periodogram used to compute the 
detection limits. 

We therefore developed a criteria representing this ratio, using two quantities: 
\begin{itemize}
\item{We first consider the period range for which the rms RV leads to some power observable in the periodogram computed over the period range 
0.01-1000 days: this period range, shown in Table~\ref{tabcrit}, is derived from a smoothed periodogram of the star RV divided by a smooth periodogram 
of the temporal sampling. For each star (except \object{HD060532} and \object{HD124850}) we compute the maximum of the periodogram (second column of Fig.\ref{res_a}) 
in that period domain. This power corresponds to pulsations for two stars, planets for one star, and a probable rotation modulation for the others. 
The signal is more uncertain for \object{HD199260}, \object{HD210302}, and \object{HD219482}, which exhibit a small RV rms. }
\item{To evaluate the relevant power at the planet period, we use the threshold defined in the periodogram when computing the LPA detection limit, 
and then average it over the periods, for the considered star. }
\end{itemize}

The ratio of the two gives the criteria $RP_1$. The same ratio can be computed by estimating the power due to the rms RV in the 2-1000 day 
periodogram (using the same period domain as above): this allows us to study \object{HD60532} (for which the power was not visible on the first periodogram), 
but eliminates the two pulsating stars from the comparison. It leads to the ratio $RP_2$. Both ratios are shown in Table~\ref{tabcrit}. 

The LPA and correlation-based methods provide the best improvement on the detection limits, and we compare in more detail 
the results obtained with these methods with this new criteria. 
Fig.\ref{crit} shows $RL_{\rm LPA}$ versus $RP_1$ and $RP_2$, respectively. We observe a very good 
correlation, showing that the larger the power due to the observed rms RV (compared to the power in the periodogram), the better the 
improvement. The discussion is similar for 
the correlation-based method. Let us consider the case of \object{$\beta$ Pic}. Because the pulsation signal dominates the periodogram, a planet at a certain 
period (in the range we consider) producing the same rms RV will produce a peak that is much stronger than the observed power, as well as some significant power 
outside this peak. The planet+star periodogram will then be dominated by the planet, leading to a strong correlation, above the threshold. On the 
other hand, if there is as much power at the period we consider as in the domain corresponding to the rms RV, the signal of a planet with a mass 
corresponding to the rms detection limit will lead to a power that will not dominate the star signal. It should be emphasized here that the 
correlation-based method is related to the pattern (hence the importance of the threshold) but also the overall amplitude of the periodogram 
(regardless of whether it dominates the star power). Fig.~\ref{seuil} (lower panel) shows the ratio of the correlation-based to the LPA detection 
limits versus the correction threshold:  there is a clear trend showing that for low thresholds the ratio is around 1 (i.e. both methods 
perform similarly), while the ratio is larger for higher thresholds (i.e. the correlation-based method does not then perform as well). 
This poorer performance of the correlation-based method therefore occurs in specific cases for which the threshold on the correlation 
(see Sect.~4.1) is large.

\subsection{Comparison of the detection limits with the different methods}

Among the methods studied in this paper, the LPA method  gives the lowest detection limits in all cases, while the rms method usually 
gives the highest, although the ratio varies significantly from one star to the other. The detection limits obtained with the three new methods are 
well-correlated with the rms detection limits, the detection limits themselves usually ranging from the rms detection limit to
a level ten times lower. Among the three tested methods, our poorest result is for the peak method (with detection limits lower than the rms detection limits for 76\% of the 
points, not counting the irrelevant ones mentioned in Sect.~5.1), and the best is for the LPA method (with detection limits lower than the rms detection limits for 96 \% of the 
points). The correlation-based method improves the detection limit in 87\% of the cases. 
The correlation and LPA methods give the best agreement between detection limits. These two methods always provide on average a lower detection limit 
than the rms method, 
and for most periods when looking at individual cases. The LPA method in general gives better detection limits, but the correlation method still 
provides better detection limits for a significant fraction of the points (23\%). The improvement with respect to the rms method  for both methods 
is clearly illustrated by the ratio of the power at the origin of the rms RV to the power in the periodogram at the period we consider. The 
differences between the correlation-based and LPA methods can also be attributed to the values of the correlation threshold, which can be large in 
some cases when the correlation-based method is then less efficient. 

Finally, we compare our results with the detection limits obtained with the bootstrap method described in \cite{zechmeister09}, which are shown in Fig.~\ref{res_a}. 
As for the rms method, the bootstrap method does not take into account the temporal structure of the power at the period we consider. If significant 
structure is observed at period $P$ (due to for example rotation modulation), the false alarm probability (hereafter fap) derived from the randomization of 
the observation will be below that power and lead to a low detection limit, while the presence of that power should prevent the detection of 
a planet with such a mass. We therefore expect the bootstrap method to underestimate the detection limit in some cases. On the other hand, since that method 
uses the maximum of the periodogram (after randomization of the signal) over the whole range of periods over which it is computed, it may in some 
cases overestimate the fap if a strong power due to the temporal sampling is observed at a specific period. Its behavior is therefore difficult to 
interpret. We find that for six of our stars there is  good agreement in general with our results (with a higher detection limit for one star), but for 
four of these stars (\object{HD124850}, \object{HD199260}, \object{HD210302}, and \object{$\beta$ Pic}), for all or most periods, the bootstrap method significantly underestimates the detection 
limits in some cases. In the \object{$\beta$ Pic} case for example, the fap is smaller than most peaks, which leads to a very low detection limit.

\subsection{Discussion of individual stars}


\subsubsection{Comparisons between stars}

Another way to assess each method is to compare our results for pairs of stars. For example, \object{HD172555} and \object{$\beta$ Pic} are two pulsating stars, with observed 
rms RV in the same category. The improvement is the largest for \object{$\beta$ Pic}, while \object{HD172555} shows a small to moderate improvement. This can be explained 
by the pulsations dominating the periodogram, which it is not the case for \object{HD172555}, probably owing to the temporal sampling. 
\object{HD115892} and \object{HD124850} have similar values of RV rms (the latter being due to a long-term trend), and the improvement is large for \object{HD115892} but small for \object{HD124850}. 
Both \object{HD60532} and \object{HD219482} have a small rms RV and observation length, but their numbers of observations are very different. The first one also has two planets. 
\object{HD60532} and \object{HD219482} lead however to similar improvements. 
\object{HD199260} and \object{HD219482} also have a close rms RV, but \object{HD199260} have twice as many points and a slightly longer observation duration. On the other hand, the sampling of \object{HD219482} appears to be quite regular. \object{HD199260} and \object{HD219482} also lead to a similar improvement. 

\subsubsection{The \object{HD10180} detection limits} 

A notable case in our sample is \object{HD10180}, for which as up to 7 planets may be present \cite[][]{lovis11a}. If we consider the two planets providing the largest 
RV signal, we find that the most massive (25.4 M$_{\rm Earth}$, i.e. 0.08 Mjup), at a period of 49.7 days, produces a small bump in the detection limits of 
Fig.~\ref{res_a}, so that the planet mass is slightly below the detection limit. The bump for the planet at 5.8 days (which is a factor of two less massive) 
is hardly visible with the correlation-based method but is observed for the LPA method. Furthermore, the planet mass is slightly lower than the LPA detection 
limit at that period. The two planet masses are therefore either very close to or slightly below the detection limits that we have determined.

\subsubsection{The \object{HD60532} detection limits}

Two planets were similarly detected around \object{HD60532} by \cite{desort08}, 
at 0.76 AU (P=204 d) and 1.58 AU (P=601 d) with masses of 1 and 2.5 Mjup, respectively. 
Dynamical studies by \cite{laskar09} found similar distances for masses 
of 3.1 and 7.4 Mjup at periods of 201.83 d and 607.06 d, respectively. With the correlation-based method, the detection limits at periods 200 and 600 days 
are, respectively, 0.9 and 4.4~Mjup. They are, respectively, 1.4 and 3.0 Mjup with the peak method, 0.9 and 2.8 Mjup with the LPA method, and 1.6 and 3.1 
with the rms method.
By definition, the detection limit computations always rely on the 
assumption that the observed signal contain no planet signal. 
Furthermore, it is always conservative: we wish all phases to give a signal 
respecting the considered criterion, although specific phases may give a detectable signal 
for masses well below the detection limit, while other phases may not. The masses determined by \cite{desort08} are very close to the detection limit 
for the planet at 204d and slightly below the detection limit for the 601d planet. This is similar to the results obtained for \object{HD10180}.

\subsubsection{The \object{HD172555} detection limits} 
 
For \object{HD172555}, \cite{quanz11} determined detection limits from imagery data obtained with NACO, i.e. at larger distances than those considered here. 
They found a detection limit of  2-3 Mjup at 15--29 AU, and 4 Mjup at 11 AU. 
In this work, in a complementary period domain (between 4 and 900 day, i.e. 0.6--2.1 AU), we found detection limits of between $\sim$4 Mjup 
(for the smallest periods) and $\sim$30--70Mjup (for the longest periods). These detection limits are significantly higher than those obtained 
for \object{$\beta$ Pic}, whose temporal sampling is much more suitable (better sampling of the pulsations, more observations). 
This example again shows the advantages  of combining RV and imaging to explore the (mass, period) domain as widely as possible.

\subsubsection{The \object{HD124850} detection limits} 

\object{HD124850} follows a trend, which impacts the detection limit computed using the rms method. We recomputed the detection limits after removing 
this trend. The rms RV becomes 31.1 m/s instead of 88.4 m/s in the original time series.  
The correlation limits are all significantly lower, by a factor of 6.9 for the correlation-based method, 3.5 for the LPA method, and 2.8 for the 
rms method. The $RL$ criteria are different from those presented in Table~\ref{tabcrit}, with values of 2.04 for $RL_{\rm corr}$ (instead of 1.07) 
and 2.62 for $RL_{\rm LPA}$ (instead of 2.21). The $RL$ values are therefore no closer to 1 because the correlation-based and LPA detection 
limits have decreased as well, as the presence of the trend (and therefore its removal) significantly impacts the periodogram both in shape and power level.

We note that the trend could be due to the presence of a planet, but the sampling is not fine enough to determine its possible period. For example, a 
planet at the LPA detection limit, in the period range studied in this paper, could produce a slope similar to the observed slope of -0.12 m/s/d 
or lower for a significant number of phases, for up to 30--40\% of the phases for some periods. The trend is therefore not necessarily caused by a long-period planet. 

\subsubsection{The \object{HD105690} detection limits}

\object{HD105690} has a strong modulation of around 4 days, which impacts the detection limits determined using the rms method. It would be useful to 
subtract this component before computing the detection limits, but this is beyond the scope of the present paper.

\subsubsection{The \object{$\beta$ Pic} detection limits} 

\object{$\beta$ Pic}  is the star for which our newly presented methods provide the greatest improvement compared to the rms method, owing to its strong pulsating 
signal and a very fine temporal sampling. This star was studied in detail in paper I.

\section{Conclusion}

We have determined and discussed robust 
detection limits in the range 4--800 days for ten stars, including \object{$\beta$ Pic}, using several new methods (Fig.~\ref{res_a}).  

We have compared the obtained detection limits, as well as the robustness, of three new methods (two of which were introduced in paper I). These three 
methods have the advantage of taking into account the temporal distribution of the power in the observed RV, owing to both the temporal sampling and 
the presence of power of various origins in the stellar signal, and not only the temporal sampling as is done in bootstrap methods such as either 
the one described in \cite{zechmeister09} and \cite{wittenmyer06} or the rms method.  
The three methods tested in this paper are sensitive to different aspects of the temporal distribution of the power and temporal sampling. 
The correlation-based detection limit is related to the pattern introduced into the periodogram by the presence of planets (with power being introduced at 
a given period and interacting with the temporal sampling), but also to the global power in the periodograms. The peak detection
limit results from the comparison between the amplitude of the peak that corresponds to the planet closest to the planet period, 
and the amplitudes of other peaks. Finally, the LPA method is related to how the total power
due to the planet in a given period range compares to the same power in the stellar periodogram. We compared these methods to both the rms-based method,
which is computed 
in the SAFIR program \cite[][]{galland05,lagrange12}, and the bootstrap method described in \cite{zechmeister09}, as this method is widely used. 

The correlation-based and LPA methods give the lowest detection limits in all cases (the latter often being better), while the rms method usually 
gives the largest, although the ratio varies significantly from one star to another. The peak method is not as efficient as the other two.  
For this small sample, the improvement with respect to the rms method  for both methods 
is clearly illustrated by the ratio of the power at the origin of the rms RV to the power in the periodogram at the period we consider. 

The correlation-based and LPA methods are also the two most robust of the three. 
The correlation-based method uses a small number of parameters and is very robust, although it is not a practical method for actually detecting 
planets in an observed time series. It is indeed difficult to determine the threshold in the correlation between the periodograms and the impact 
of the presence of several planets on the pattern observed in the periodogram. The LPA method is also very robust, although the period window 
has a significant impact on the result. The peak method however is not as robust, as in some cases it is impossible to identify a planet peak in 
the periodogram owing to the temporal sampling of the observations, even for a very high planet mass. It also relies on many parameters. The main 
limitation of the peak method, despite it being the closest to the method actually used to detect a planet on a RV time series, is that 
one needs to identify automatically the peak corresponding to the planet period: in a significant number of cases, the planet peak is indeed not the 
largest one, even for a very massive planet, owing to the temporal sampling. In addition, the study of the robustness provides an estimation of 
the uncertainty in the derived detection limits. 

We conclude that the rms method is ideal for achieving a quick look. Both this method and the bootstrap method provides an efficient determination of the 
detection limits. However, we point out that, 
at least for the stars we have studied (F-G stars, with stellar activity and / or planets, MS stars), these methods may not give the best results, 
as they do not take into account the temporal response of the RV signal (due to either, for example, stellar activity or the presence of planets).
Therefore, to obtain a more robust estimate, we recommand using both the correlation-based method and the LPA method, especially for times series 
corresponding to many observations that have well-defined peaks in the temporal periodogram. 
The use of more than one method is also useful for estimating the uncertainty in the detection limits. 
This study is in principle limited by the size of the sample and selection effects. However, even for a small number of stars that cover a large range of parameters, 
it has allowed us to derive some clear indications of to the expected improvement. An improvement exists even for stars with a very low RV jitter, so that the whole sample exhibits a coherent behavior.

\begin{acknowledgements}
We acknowledge financial support from the French Programme de Plan\'etologie (PNP, INSU). 
We also acknowledge support from the French National Agency (ANR) through project grant ANR10-BLANC0504-01.
\end{acknowledgements}

\bibliographystyle{aa}
\bibliography{bib19163}

\listofobjects

\end{document}